\documentclass[aps,pre,twocolumn,superscriptaddress]{revtex4}
\usepackage{graphics}
\usepackage{epsfig}
\usepackage{amsmath}
\usepackage{amsthm}
\usepackage{amsfonts}
\usepackage{subfigure}
\usepackage{graphicx}
\usepackage{mathrsfs}
\usepackage{revsymb}
\usepackage{color}
\newcommand{\ket}[1]{\vert #1 \rangle}

\newcommand{\meanvalue}[3]{\langle #1 \vert #2 \vert #3 \rangle}
\newcommand{\ketbra}[2]{\vert #1 \rangle \langle #2 \vert}
\newcommand{\imm}{{\rm i }}
\newcommand{\bea}{\begin{eqnarray}}   
\newcommand{\eea}{\end{eqnarray}}

\begin{document}
\title{Dicke coupling by feasible local measurements at 
the superradiant quantum phase transition}
\author{M. Bina}
\affiliation{Dipartimento di Fisica, Universit\`a degli 
Studi di Milano, I-20133 Milano, Italy}
\author{I. Amelio}
\affiliation{Dipartimento di Fisica, Universit\`a degli 
Studi di Milano, I-20133 Milano, Italy}
\author{M. G. A. Paris}
\email{matteo.paris@fisica.unimi.it}
\affiliation{Dipartimento di Fisica, Universit\`a degli 
Studi di Milano, I-20133 Milano, Italy}
\affiliation{CNISM, UdR Milano Statale, I-20133 Milano, Italy}
\affiliation{INFN, Sezione di Milano, I-20133 Milano, Italy}
%%%%
\date{\today}
%%%%
\begin{abstract}
We address characterization of many-body superradiant systems and
establish a fundamental connection between quantum criticality and the
possibility of {\em locally} estimating the coupling constant, i.e
extracting its value by probing only a portion of the whole system.  In
particular, we consider Dicke-like superradiant systems made of an
ensmble of two-level atoms interacting with a single-mode radiation
field at zero effective temperature, and address estimation of the
coupling by measurements performed only on radiation.  At first, we
obtain analytically the Quantum Fisher Information (QFI) and show that
optimal estimation of the coupling may be achieved by tuning the frequency
of the radiation field to drive the system towards criticality. The
scaling behavior of the QFI at the critical point is obtained explicitly upon
exploiting the symplectic formalism for Gaussian states.  We then
analyze the performances of feasible detection schemes performed only on
the radiation subsystem, namely homodyne detection and photon counting,
and show that the corresponding Fisher Informations (FIs) approach the
global QFI in the critical region.  We thus conclude that criticality is
a twofold resource. On the one hand, global QFI diverges at the critical point,
i.e. the coupling may be estimated with the arbitrary precision. On the
other hand, the FIs of feasible local measurements, (which are generally
smaller than the QFI out of the critical region), show the same scaling
of the global QFI, i.e. optimal estimation of coupling may be achieved
by locally probing the system, despite its strongly interacting nature. 
\end{abstract} 
\maketitle
\section{Introduction}
Quantum phase transitions (QPTs) occur at zero temperature and 
demarcate two statistically distinguishable ground states corresponding to
different quantum phases of the system \cite{QPT}. In the proximity of
the critical point, small variations of a parameter driving the QPT
cause abrupt changes in the ground state of the system. Criticality is,
thus, a resource for precision measurements \cite{Fidelity} since 
driving the system to the critical region makes it extremely sensitive 
to perturbations, either affecting an internal parameter such as its 
coupling constant, or due to fluctuations of environmental parameters, 
e.g. temperature fluctuations. 
\par
It is often the case that those parameters are not directly measurable. 
In these cases, the determination of their values should be pursued 
exploiting indirect observations and the technique of parameter estimation. In 
this situations, the maximum
information extractable from an indirect estimation of the parameters 
is the so-called Fisher Information (FI), which itself determines the best 
precision of the estimation strategy via the Cramer-Rao theorem
\cite{Cramer}. Upon optimizing over all the possible quantum measurements 
one obtains the Quantum Fisher Information (QFI), which  depends
only on the family of states (density operators) describing the ground
state of the considered system \cite{Helstrom,ParisQET} as a function of 
the parameter of interest. In turn, the QFI sets the ultimate quantum bound 
to precision for any inference strategy aimed at estimating a given 
parameter. 
\par
In the recent years, the connection between quantum criticality and 
parameter estimation has been addressed from differente perspectives
\cite{Zanardi,CozziniGiorda,ZanardiGiorda,Invernizzi,Garnerone,InvernizziParis,Mandarino}, showing that the QFI is indeed (substantially) enhanced in 
correspondence of the critical point \cite{Fidelity}. The
fundamental interpretation of this relationship lies in the geometrical
theory of quantum estimation, for which Hilbert distances between states
are translated into modifications of the physical parameters
\cite{Zanardi, CozziniGiorda, ZanardiGiorda}. Criticality as a resource
for quantum metrology has been investigated in several critical systems
\cite{Invernizzi, Garnerone, InvernizziParis, Mandarino}. Nonetheless,
finding an optimal observable which also corresponds to a feasible
detection scheme is usually challenging, especially for strongly
interacting systems where the entangled nature of the ground state 
usually leads to an inseparable optimal observable.
\par
In this paper we consider the superradiant QPT occurring in the 
Dicke model, which describes the strong interaction of a single-mode
electromagnetic field and an ensemble of two-level atoms \cite{Dicke}.
The radiation mode in the superradiant phase acquires macroscopic
occupation as a consequence of cooperative excitation of the atoms 
in the strong coupling regime. The Dicke QPT has been extensively studied
in the past years considering also generalizations of the original work
of Dicke \cite{WangHioe, Hioe, HeppLieb}, or focussing on the
quantum-cahotic properties of the system \cite{EmaryBrandesPRL,
EmaryBrandesPRE}. Recent theoretical studies concerning entanglement and
squeezing of the Dicke QPT have been carried on \cite{Nataf}, also in
relation to the QFI of radiation and atomic subsystems separately
\cite{Nori}. Some implementations in cavity-QED \cite{Carmichael} and
circuit-QED \cite{Viehmann, Solano} systems, together with
computing applications via multimodal disordered couplings
\cite{Rotondo}, have been proposed. Eventually, recent experimental
realizations of the Dicke QPT involving Bose-Einstein condensates in
optical cavities \cite{Esslinger}, cavity-assisted Raman transitions
with $\text{Rb}^{87}$ atoms \cite{Barrett} or NV-centers in diamond coupled
to superconducting microwave cavities \cite{Rabl}, have been performed.
\par
Motivated by the renewed experimental and theoretical interests in the
Dicke QPT, we address the characterization of its coupling constant
and analyze in details whether optimal estimation is possible using 
only feasible local measurements, i.e. whether the ultimate 
precision allowed by quantum mechanics may be achieved 
by probing only a portion of the whole system.  
\par
The paper is structured as follows. In Sec. \ref{s:QET} we introduce 
the properties of Gaussian states and symplectic transformations, 
together with some elements of quantum
estimation theory (QET) in the Gaussian continuous-variable formalism.
In Sec. \ref{s:Dicke}, we briefly review the Dicke model at zero
temperature, establish notation and find the Gaussian ground states of 
the two phases of the system. In Sec. \ref{s:QFI} we evaluate the QFI as 
a function of the radiation-atoms coupling parameter and discuss its 
properties. Eventually, in Sec. \ref{s:FI} we
present our main results concerning the analysis of the FI associated to
two locally feasible observables, homodyne detection and photon
counting. We will show that these feasible measurements
allow to achieve optimal estimation of the coupling parameter by probing
only the radiation part of the system. 

%%%%%%%%%%%%%%%%%%
%%%%%%%%%%%%%%%%%%
\section{Tools of Quantum estimation theory for Gaussian states}
\label{s:QET}
In this section we briefly introduce the formalism of Gaussian states for continuous-variable bosonic systems and of symplectic diagonalization of quadratic Hamiltonians \cite{GSFerraro,GSBraunstein,GSOlivares}. 
%At a later stage, after introducing the Dicke model, we address the underlying QPT expressing the ground states, relatively to the two phases, in terms of Gaussian states.

%%%%%%%%%%%%%%%%%%

\subsection{Gaussian states and symplectic transformations}\label{Gaussian}

A system composed by $M$ bosonic modes is described by quantized fields $\hat{a}_m$ satisfying the commutation relation $[\hat{a}_m,\hat{a}_l^\dag]=\delta_{m,l}$. An equivalent description is provided, through the Cartesian decomposition of field modes, in terms of position- and momentum-like operators $\hat{x}_m=(\hat{a}_m+\hat{a}_m^\dag )/\sqrt{2}$ and $\hat{p}_m=\imm(\hat{a}_m^\dag-\hat{a}_m )/\sqrt{2}$. Introducing the vector of ordered quadratures $\vec{R}=(\hat{x}_1,\hat{p}_1,\cdots,\hat{x}_M, \hat{p}_M)^T$ and the symplectic matrix
\begin{equation}
\Omega=\bigoplus_{m=1}^M\omega_m,\quad \omega_m=\begin{pmatrix} 
0 & 1\\
-1 & 0
\end{pmatrix},
\end{equation}
the commutation relations become $[R_i,R_j]=\imm\Omega_{ij}$. The state $\hat{\varrho}$ of a system of $M$ bosonic modes can be described in the phase space by means of the characteristic function, defined as $\chi[\hat{\varrho}](\vec{\alpha}) \equiv {\rm Tr}[\hat{\varrho} \hat{D}(\vec{\alpha})]$, where $\hat{D}(\vec{\alpha})=\bigotimes_{m=1}^M\exp\{\alpha_m \hat{a}_m^\dag-\alpha_m^* \hat{a}_m\}$ is the displacement operator and $\vec{\alpha}=\{\alpha_1,\ldots,\alpha_M\}$, with complex coefficients $\alpha_m=(\alpha^{(r)}_m+\imm\, \alpha^{(i)}_m)/\sqrt{2}$ and $\{\alpha^{(r)}_m,\alpha^{(i)}_m\}\in \mathbb{R}$. It is responsible for rigid translations of states in the phase space, allowing to express any coherent state as a displaced vacuum state $\ket{\vec{\alpha}}=\hat{D}(\vec{\alpha})\ket{0}$. Equivalently, in the cartesian representation, the displacement operator can be written in the compact form $\hat{D}(\vec{\Lambda})= \exp\{ -\imm \vec{\Lambda}^T \Omega \vec{R} \}$, with $\vec{\Lambda}=\{ \alpha^{(r)}_1,\alpha^{(i)}_1,\ldots,\alpha^{(r)}_M,\alpha^{(i)}_M \}$, acting on the vector of quadratures as $\hat{D}^\dag(\vec{\Lambda}) \vec{R}\, \hat{D}(\vec{\Lambda})=\vec{R}+\vec{\Lambda}$.
\par
A density operator $\hat{\varrho}$ describing the state of a system of $M$ bosonic modes, is called Gaussian when its characteristic function $\chi[\hat{\varrho}](\vec{\Lambda}) \equiv Tr[\hat{\varrho} \hat{D}(\vec{\Lambda})]$ is Gaussian in the cartesian coordinates $\vec{\Lambda}$ and reads
\begin{equation}
\chi \left[ \hat{\varrho} \right] (\vec{\Lambda}) = \exp \left \{ -\frac{1}{2} \vec{\Lambda}^T \Omega \sigma \Omega^T \vec{\Lambda} -\imm \vec{\Lambda}^T \Omega \langle \vec{R} \rangle \right \}, 
\label{Gaussdef}
\end{equation}
or, equivalently, when the associated Wigner function has the Gaussian form
\begin{equation}\label{Wigner}
W[\hat{\varrho}](\vec{X})=\frac{\exp \left\{ -\frac{1}{2} (\vec{X} - \langle \vec{R} \rangle)^T\sigma^{-1}(\vec{X}-\langle \vec{R} \rangle) \right\}}{\pi^M\sqrt{\text{Det}[\sigma]}},
\end{equation}
the two being related by the Fourier transform 
\begin{equation}
W[\hat{\varrho}](\vec{X})=\frac{1}{(2\pi)^{2 M}}\int d^{2M}\vec{\Lambda}\exp \{\imm \,\vec{\Lambda}^T\Omega\vec{X} \} \chi \left[ \hat{\varrho} \right] (\vec{\Lambda}).
\end{equation}
A Gaussian state is completely determined by the first-moments vector $ \langle \vec{R}\, \rangle$ and the second moments encoded in the covariance matrix (CM) $\sigma$, of elements
\begin{equation}
\sigma_{ij} = \frac{1}{2} \langle R_i R_j + R_j R_i \rangle - \langle R_i  \rangle \langle R_j \rangle,
\end{equation}
which allows to write the Heisenberg uncertainty relation as $\sigma+\frac{\imm}{2}\Omega\geq 0$. The purity $\mu=Tr[\hat{\varrho}^2]$ of a Gaussian state is expressed in terms of the CM by the relation $\mu=( 2^M \sqrt{\text{Det}[\sigma]}\,)^{-1}$.
\par
A property of Gaussian states, which will reveal to be useful in the following calculations, is that the reduced density matrix, obtained by means of the partial trace operation over the degrees of freedom of a subsystem, keeps its Gaussian character \cite{Adesso}. For instance, exploiting the Glauber representation of a density operator of a bipartite state, with cartesian coordinates $\vec{\Lambda}=(\vec{\Lambda}_{a_1},\vec{\Lambda}_{a_2})$
\begin{equation}
\hat{\varrho}_{a_1a_2}=\frac{1}{(2\pi)^2}\int_{\mathbb{R}^{4}}\text{d}^{4}\vec{\Lambda}\;\chi[\hat{\varrho}_{a_1a_2}](\vec{\Lambda})\hat{D}^\dag (\vec{\Lambda}),
\end{equation}
together with $\text{Tr}[\hat{D}(\vec{\Lambda}_{a_1})]=(2\pi)\delta^{(2)}(\vec{\Lambda}_{a_1})$, then the reduced density operator $\hat{\varrho}_{a_1}$ is a Gaussian state with an associated characteristic function $\chi[\hat{\varrho}_{a_1a_2}](\vec{\Lambda}_{a_1},0)=\text{Tr}[\hat{\varrho}_{a_1} \hat{D}(\vec{\Lambda}_{a_1})]$.
\par
To become more familiar with these concepts, we list here some examples of Gaussian states. A single-mode system in a equilibrium with a thermal environment is described by the density operator $\hat{\nu}_\text{th}(\bar{n})=\sum_{k=0}^\infty\bar{n}^k(1+\bar{n})^{-(1+k)}\ketbra{k}{k}$ expressed on the Fock basis $\{\ket{k}\}_0^\infty$. The corresponding covariance matrix is $\sigma_\text{th}=(1+\bar{n})/2$, with $\bar{n}$ the number of average thermal photons. Other examples include the classes of coherent states and squeezed states, for which the uncertainty relation $\sigma+\frac{\imm}{2}\Omega\geq 0$ is saturated with $\langle \Delta \hat{x} ^2 \rangle \langle\Delta \hat{p}^2 \rangle=1/4$. All coherent states have $\langle \Delta \hat{x}^2 \rangle=\langle \Delta \hat{p}^2 \rangle=1/2$, whereas squeezed states possess a covariance matrix of the kind $\sigma_\text{sq}=\frac12\text{Diag}({\rm e}^{2 r},{\rm e}^{-2r})$, where $\langle \Delta \hat{x} ^2 \rangle \neq \langle \Delta \hat{p}^2 \rangle$ and $r\in\mathbb{R}$ is a real squeezing parameter. A generic squeezed state is obtained from the vacuum by applying the unitary operator $\hat{S}(\xi)=\exp\{(\xi (\hat{a}^\dag)^2-\xi^* \hat{a}^2)/2\}$, with complex squeezing parameter $\xi=r{\rm e}^{\imm\psi}$. The most general single-mode Gaussian state is a displaced squeezed thermal state (DSTS) described by the density operator $\hat{\varrho}=\hat{D}(\gamma)\hat{S}(r) \hat{\nu}_\text{th}(\bar{n}) \hat{S}^\dag (r)\hat{D}^\dag (\gamma)$. For two-mode systems such a general form does not exist, but a relevant subclass of bipartite Gaussian states is given by the squeezed thermal states $\hat{\varrho}_{a_1 a_2}=\hat{S}_2(\xi)\hat{\nu}_\text{th}(\bar{n}_1)\otimes \hat{\nu}_\text{th}(\bar{n}_2)\hat{S}_2^\dag (\xi)$, where $\hat{S}_2(\xi)=\exp\{\xi \hat{a}_1^\dag \hat{a}_2^\dag -\xi^* \hat{a}_1 \hat{a}_2\}$ is the two-mode squeezing operator. \\
An important property of Gaussian states is related to transformations induced by quadratic Hamiltonians.
Gaussian states preserve their Gaussian character under symplectic transformations of coordinates $\vec{R}\to F \vec{R}+\vec{d}$, where $\vec{d}$ is a vector of real numbers, leaving unchanged the Hamilton equations of motion and fulfilling the symplectic condition $F\Omega F^T=\Omega$. Thus, the first-moment vector and the CM of a Gaussian state follow the transformation rules
\begin{equation}\label{SimpTrans}
 \langle \vec{R} \rangle \to F \langle \vec{R} \rangle + \vec{d} \;, \quad \sigma\to F\sigma F^T.
\end{equation}
Moreover, symplectic transformations originate from Hamiltonians at most bilinear in the field modes (quadratic) and the diagonalization process of these Hamiltonians goes under the name of symplectic diagonalization, which transforms the coordinates by preserving canonical commutation relations. Symplectic transformations possess the property of unitary determinant $\text{Det}[F]=1$. As an example, consider a thermal state $\hat{\nu}_\text{th}(\bar{n})$ evolving under the single-mode real squeezer $\hat{S}(r)$. The associated symplectic matrix is $F=\text{Diag}({\rm e}^r,{\rm e}^{-r})$ and, according to Eq. (\ref{SimpTrans}), the CM transforms as $\sigma=\frac12(1+2\bar{n})\text{Diag}({\rm e}^{2r},{\rm e}^{-2r})$, which is the CM of a squeezed thermal state.

In the light of the properties of symplectic transformations and writing the CM of Gaussian bipartite states in the most general way as $\sigma=\begin{pmatrix} 
A  & C\\
C^T & B
\end{pmatrix},$ it is possible to identify four symplectic invariants given by $I_1=\text{Det}[A]$, $I_2=\text{Det}[B]$, $I_3=\text{Det}[C]$ and $I_4=\text{Det}[\sigma]$. The symplectic eigenvalues of a CM can be expressed in terms of these invariants as
\begin{equation}\label{SympEig}
d_\pm=\sqrt{\frac{I_1+I_2+2I_3\pm\sqrt{(I_1+I_2+2I_3)^2-4 I_4}}{2}},
\end{equation}
from which we can straightforwardly rewrite the uncertainty relation as $d_-\geq 1/2$. Pure Gaussian states have $I_4=1/16$ and $I_1+I_2+2I_3=1/2$. The separability of the two subsystems is formalized in terms of the criterion of positivity under partial transpose (ppt) \cite{Peres_ppt}, which can be written in terms of the symplectic invariants as $\tilde{d}_- \geq 1/2$, where
\begin{equation}\label{SympEigPT}
\tilde{d}_\pm=\sqrt{\frac{I_1+I_2-2I_3\pm\sqrt{(I_1+I_2-2I_3)^2-4 I_4}}{2}}
\end{equation}
are the symplectic eigenvalues of the CM of the partially transposed density operator describing a bipartite Gaussian state. A measure of entanglement is, thus, provided by the logarithmic negativity \cite{Vidal_LogNeg}
\begin{equation}\label{LN}
E_{\mathcal{N}}(\sigma)=\text{max}\{0,-\ln 2\tilde{d}_-\},
\end{equation}
which quantifies monotonically the amount of violation of the ppt-criterion.
%%%%%%%%%%%%%%%%%%
\subsection{Local QET}
Whenever a parameter of a physical system is not directly accessible by an observable, it is always possible to infer its average value by means of classical estimation theory inspecting the set of data $\{x\}$ of an indirect measurement. 
Let us suppose that an observable $\hat{\mathcal{X}}$ is measured on the considered physical system described by a parameter-dependent density operator $\hat{\varrho}_\lambda$. A set of data $\{x_1,\ldots, x_m \}$, corresponding to the possible outcomes of $\hat{\mathcal{X}}$, is then collected according to the distribution $p(x | \lambda)=\text{Tr} [\hat{\varrho}_\lambda \hat{\mathcal{X}} ]$ provided by the Born rule, which describes the conditional probability to obtain an outcome $x$ given the value of the parameter $\lambda$. The value of the parameter $\lambda$ is then  inferred from the statistics of an estimator $\bar{\lambda}=\bar{\lambda} (x_1,\ldots,x_m)$, evaluating its average value $E[\bar{\lambda}]$ and variance $\text{Var}_{\lambda} =E[\bar{\lambda}^2]-E[\bar{\lambda}]^2$ (valid for any unbiased estimator $E[\bar{\lambda}]=\lambda$). From classical estimation theory, optimal estimators saturate the Cram\'er-Rao bound
\begin{equation}\label{CR}
\text{Var}_{\lambda} \geq \frac{1}{m F_{\hat{\mathcal{X}}}(\lambda)},
\end{equation}
where the FI $F_{\hat{\mathcal{X}}}(\lambda)$ is the maximum information extractable from a measurement of the observable $\hat{\mathcal{X}}$ and reads
\begin{equation}\label{FI}
F_{\hat{\mathcal{X}}}(\lambda)=\int_{\mathbb{R}} \text{d} x\, p(x | \lambda)\, \left ( \partial_\lambda \big [ \ln p(x | \lambda)\big ] \right )^2.
\end{equation}
The ultimate limit to the precision in an estimation process is given the quantum Cram\'er-Rao bound
\begin{equation}\label{QCR}
\text{Var}_{\lambda} \geq \frac{1}{m H(\lambda)},
\end{equation}
where the QFI $H(\lambda)$ does not depend on measurements but only on the probe state $\hat{\varrho}_\lambda$. The QFI is the result of a maximization over all the possible observables on the physical system and it is such that $H(\lambda)\geq F_{\hat{\mathcal{X}}}(\lambda)$. The QFI is analytically computable as $H(\lambda)=\text{Tr}[\hat{\varrho}_\lambda \hat{\mathcal{L}}_\lambda^2]$, i.e. in terms of the hermitean operator $\hat{\mathcal{L}}_\lambda$ called symmetric logarithmic derivative (SLD), implicitly defined as
\begin{equation}\label{SLD}
\partial_\lambda \hat{\varrho}_\lambda\equiv\frac{\hat{\mathcal{L}}_\lambda \hat{\varrho}_\lambda+\hat{\varrho}_\lambda \hat{\mathcal{L}}_\lambda}{2}.
\end{equation}
The SLD operator represents the optimal positive-operator valued measurement (POVM) saturating the Cram\'er-Rao bound (\ref{QCR}). 
\par
Criticality at a QPT is a resource for quantum estimation as a small change in the parameter $\lambda$ yields a drastic change in the ground state at the boundary of the critical parameter, thus allowing the QFI to diverge. It is, thus, desirable to find an optimal observable maximizing the FI to the values of the QFI in order to achieve the best precision in the parameter estimation.
\par
In the context of Gaussian states it is possible to derive analytical expressions for the QFI and the SLD operator \cite{Jiang}, which depend on the physical parameters characterizing the state of the system. Exploiting the notions outlined in Sect. \ref{Gaussian} and redefining the partial derivation as $\partial_\lambda(f)\equiv\dot{f}$, the QFI and SLD for a generic Gaussian state read
\begin{align}
\label{QFI_Gaussian}  H(\lambda)&= \text{Tr} \left[ \Omega^T \dot{\sigma} \Omega \Phi \right] + {\langle \dot{\vec{R}}\, \rangle}^T \sigma^{-1} {\langle \dot{\vec{R}} \rangle}\\ 
\label{SLD_Gaussian} \mathcal{L}_\lambda&=\vec{R}^{\,T} \Phi \vec{R} + \vec{R}^{\,T} \vec{\zeta} - \nu,
\end{align}
where $\nu=\text{Tr}[\Omega^T\sigma\,\Omega\,\Phi]$ is related to the property of the SLD (\ref{SLD}) to have zero-mean value $\text{Tr}[\hat{\varrho}_\lambda\hat{\mathcal{L}}_\lambda]=0$. For pure Gaussian states all the quantities in Eqs. (\ref{QFI_Gaussian})-(\ref{SLD_Gaussian}) are easy to compute and read $\Phi=-\dot{\sigma}$ and $\vec{\zeta} = \Omega^T \sigma^{-1} \langle \dot{\vec{R}} \rangle $.\\
In the following we will apply these tools to the Dicke model, in order to exploit the predicted QPT, and the corresponding ground states, for the estimation of the coupling parameter $\lambda$.
\section{Dicke quantum phase transition for quantum estimation}\label{s:Dicke}
In this section we describe the Dicke QPT in the Gaussian formalism, suitable for establishing a tight connection with local estimation theory performed with measurements typical of quantum optics. In particular we derive the Gaussian ground states corresponding to the normal and superradiant phases, computing the amount of entanglement and the scaling behaviors of the associated QFI and SLD.

\subsection{The superradiant QPT}
The Dicke model \cite{Dicke} describes the interaction between a dense collection of $N$ two-level atoms (spin objects) with transition frequency $\omega_0$, assumed to be equal for all the spins, and a single radiation mode (bosonic field) of frequency $\omega$, which is characterized in terms of annihilation and creation operators, $\hat{a}_1$ and $\hat{a}_1^\dag$ respectively. The coupling between the two quantum systems is suitably described within the dipole approximation, where each atom couples to the electric field of radiation with a coupling strength $\lambda$:
\begin{equation}\label{H}
\hat{H}_{(1,2)} = \omega_{0} \hat{J}_{z} + \omega \hat{a}_1^{\dagger} \hat{a}_1 + \frac{\lambda}{\sqrt{N}} (\hat{a}_1^{\dagger} + \hat{a}_1) (\hat{J}_{+} + \hat{J}_{-}).
\end{equation}
Overall, the atomic subsystem can be described as a pseudospin of length $N/2$ by the collective spin operators $\hat{J}_z=\frac12\sum_{i=1}^N\hat{\sigma}_z^{(i)}$ and $\hat{J}_{\pm}=\sum_{i=1}^N \hat{\sigma}_\pm^{(i)}$, where $\{\hat{\sigma}_z^{(i)},\hat{\sigma}_\pm^{(i)}\}$ is the set of Pauli matrices that completely characterize single two-level systems.  
\par
The diagonalization of Hamiltonian $(\ref{H})$ is performed employing the Holstein-Primakoff (H-P) representation of the atomic spin operators \cite{H-P,Hillery}, namely $ \hat{J}_{+} = \hat{a}_2^{\dag} \sqrt{N - \hat{a}_2^{\dag} \hat{a}_2} $, $  \hat{J}_{-} =  \sqrt{N - \hat{a}_2^{\dag} \hat{a}_2} \; \hat{a}_2$ and $\hat{J}_{z} =  \hat{a}_2^{\dag}  \hat{a}_2 - \frac{N}{2} $, where $\hat{a}_2$ and $\hat{a}_2^\dag$ are bosonic fields satisfying $[\hat{a}_2, \hat{a}_2^\dagger]=1$. As will become soon clearer, the bosonic fields $\{\hat{a}_1,\hat{a}_2\}$ are allowed to have macroscopic occupations in such a way that $\hat{a}_1 \to \hat{a}_1-\alpha\sqrt{N}$ and $\hat{a}_2\to\hat{a}_2+\beta\sqrt{N}$
%$\hat{a}_1 \to \hat{a}_1'=\hat{D}_1^\dag(-\alpha\sqrt{N}) \hat{a}_1\hat{D}_1(-\alpha\sqrt{N})$ and $\hat{a}_2\to\hat{a}_2'=\hat{D}_2^\dag(-\beta\sqrt{N}) \hat{a}_2 \hat{D}_2(-\beta\sqrt{N})$, 
with $\{\alpha,\beta\}\in\mathbb{R}$. Now, we consider the thermodynamic limit, for which the ratio $N/V$ is constant as $N,V\to\infty$, being $N$ the number of atoms and $V$ the corresponding occupied volume, and expand the H-P representation keeping only the terms proportional to $\sqrt{N}$.  Applying stability considerations, for which linear terms proportional to $\sqrt{N}$ must vanish \cite{Hillery}, we obtain the expression for the displacing parameters 
%\begin{subequations}\begin{align}
%&\quad \alpha^{(n)}=0 \;, \; \beta^{(n)}=0 \quad \text{for } \lambda<\lambda_c\\
%&\quad \alpha^{(s)}=\frac{\lambda}{\omega}\sqrt{1-\frac{\lambda_c^4}{\lambda^4}} \; , \; \beta^{(s)}=\sqrt{\frac12-\frac{\lambda_c^2}{2\lambda^2} } \quad \text{for} \lambda>\lambda_c
%\end{align}\end{subequations}
\begin{equation}\label{displacement}
\left\{
\begin{array}{l}
\alpha=\pm\frac{\lambda}{\omega}\sqrt{1-k^2}\vspace{1mm} \\
\beta=\pm\sqrt{\frac{1-k}{2} }
\end{array} \right.
\end{equation}
where we introduced the dimensionless critical parameter
\begin{equation}
k\equiv\left\{
\begin{array}{l}
1 \quad \text{for } \lambda < \lambda_c \\
 \frac{\lambda_c^2}{\lambda^2}  \quad \text{for } \lambda > \lambda_c
\end{array}\right.
\end{equation}
and the critical coupling strength $\lambda_c\equiv\sqrt{\omega \,\omega_0}/2$. As it is now clear, the macroscopic occupation of the two subsystems individuates the phase transition between a \textit{normal} phase for $\lambda < \lambda_c$ and a \textit{superradiant} phase for $\lambda > \lambda_c$. The Hamiltonian of this system can be cast in diagonal form (see Ref.~\cite{EmaryBrandesPRE}), by introducing a new couple of bosonic modes $\{\hat{b}_-,\hat{b}_+\}$ which satisfy the commutation relations $[\hat{b}_-,\hat{b}_-^\dag]=[\hat{b}_+,\hat{b}_+^\dag]=1$, and describe two independent harmonic oscillators
\begin{equation}\begin{split}
\hat{H}_{(-,+)}&=\,\varepsilon_- \hat{b}_-^\dag \hat{b}_- + \varepsilon_+ \hat{b}_+^\dag \hat{b}_+ \,+\\& +
\frac12 \left( \varepsilon_- + \varepsilon_+ -\omega-\frac{\omega_0}{k} \right ) -\frac{\omega_0(1+k^2)}{2k}\frac{N}{2},
\end{split}\end{equation}
where the eigenfrequencies are given by
\begin{equation}
2 \varepsilon_\pm^2=\omega^2 + \frac{\omega_0^2}{k^2} \pm \sqrt{\left[ \frac{\omega_0^2}{k^2}-\omega^2 \right]^2 + 16 \lambda^2\, \omega\, \omega_0\, k}\;.
\end{equation}
The diagonalization in Ref. \cite{EmaryBrandesPRE} can be obtained by performing the symplectic transformation $F=F_3\circ F_2\circ F_1$:
\begin{equation}\label{SympTrans}\begin{split}
&F_1 =\text{Diag} \left (\frac{1}{\sqrt{\omega}},\sqrt{\omega},\frac{1}{\sqrt{\tilde{\omega}}},\sqrt{\tilde{\omega}}\right )\\
&F_2=\begin{pmatrix}
\cos \theta  \,\mathbb{I}_2 & -\sin \theta \,  \mathbb{I}_2\\
\sin \theta \, \mathbb{I}_2  & \cos \theta \, \mathbb{I}_2
\end{pmatrix} \\
&F_3=\text{Diag} \left ( \sqrt{\varepsilon_-},\frac{1}{\sqrt{\varepsilon_-}},\sqrt{\varepsilon_+},\frac{1}{\sqrt{\varepsilon_+}}\right ).
\end{split}\end{equation} 
Symplectic matrix $F_1$ corresponds to a local squeezing $\hat{S}_{\text{loc}}^{(1)}=\hat{S}(-\log (\sqrt{\omega}))\otimes \hat{S}(-\log (\sqrt{\tilde{\omega}}))$ applied to the quadratures of the atomic and photonic subsystems, with $\tilde{\omega}=\omega_0 (1+k)/2k$. Then the rotation $\hat{U}(\theta)=\exp\{-\imm\theta(\hat{x}_1 \hat{p}_2-\hat{x}_2 \hat{p}_1)\}$ is associated to the symplectic matrix $F_2$ ($\mathbb{I}_2$ is a $2\times 2$ identity matrix) and allows to eliminate the interaction term in the Hamiltonian upon the choice of the angle $2\theta=\tan^{-1}\big [ 4\lambda\sqrt{\omega\,\omega_0\, k}\,k^2 / (\omega_0^2-k^2\,\omega^2) \big]$. Eventually, a second local squeezing $\hat{S}_{\text{loc}}^{(2)}=\hat{S}(\log (\sqrt{\varepsilon_-})\otimes \hat{S}(-\log (\sqrt{\varepsilon_+}))$, related to the symplectic transformation $F_3$, completes the diagonalization. The ground state of the diagonalized Hamiltonian $\hat{H}_{(-,+)}$ is the vacuum state $\ket{\psi}\equiv\ket{0}_-\otimes\ket{0}_+$, with CM $\sigma_{\psi}=\mathbb{I}_4/2$. Accounting for the displacement $\hat{D}_{12}\equiv\hat{D}_1(\alpha\sqrt{N})\otimes \hat{D}_2(-\beta\sqrt{N})$ responsible for the macroscopic occupation of the original modes $\{\hat{a}_1,\hat{a}_2\}$ in the superradiant phase, the form of the Gaussian ground state $\ket{\Psi}$ is straightforwardly obtained by means of the transformation $\ket{\Psi}=\hat{D}_{12}\hat{U}_F\ket{\psi}$, where $\hat{U}_F\equiv\hat{S}_{\text{loc}}^{(1)} \, \hat{U}(\theta)\, \hat{S}_{\text{loc}}^{(2)}$ is the unitary evolution of the modes associated to the symplectic transformation $F$. The corresponding CM $\sigma\equiv\sigma_\Psi$ and first-moment vector $\langle\vec{R}\rangle$ are derived using Eqs. (\ref{SimpTrans}):
\begin{align}
\sigma&=F\sigma_{\psi}F^T=\begin{pmatrix}
\sigma_{11} & 0 & \sigma_{13} &0\\
0 & \sigma_{22} & 0 & \sigma_{24} \\
\sigma_{31} & 0 & \sigma_{33} & 0\\
0 & \sigma_{42} & 0 & \sigma_{44}
\end{pmatrix} \label{CM}\\[3mm]
\langle\vec{R}\rangle &= (\alpha\sqrt{2 N},0,-\beta\sqrt{2 N},0)^T, \label{R}
\end{align}
where
\begin{align}
\sigma_{11}&=\frac{\omega}{2}\left ( \frac{\cos^2\theta}{\varepsilon_-}+\frac{\sin^2\theta}{\varepsilon_+} \right )\nonumber \\
\sigma_{22}&=\frac{1}{2\omega} \left ( \varepsilon_- \cos^2\theta+\varepsilon_+ \sin^2\theta \right ) \nonumber \\
\sigma_{33}&=\frac{\tilde{\omega}}{2}\left ( \frac{\cos^2\theta}{\varepsilon_+}+\frac{\sin^2\theta}{\varepsilon_-} \right ) \\
\sigma_{44}&=\frac{1}{2\tilde{\omega}} \left ( \varepsilon_+ \cos^2\theta+\varepsilon_- \sin^2\theta \right ) \nonumber \\
\sigma_{13}&=\sigma_{31}=\frac{\sqrt{\omega\,\tilde{\omega}}\sin 2\theta}{4} \left( \frac{1}{\varepsilon_+} - \frac{1}{\varepsilon_-} \right)\nonumber \\
\sigma_{24}&=\sigma_{42}=-\frac{\sin 2\theta}{4\sqrt{\omega\,\tilde{\omega}}}\left ( \varepsilon_- -\varepsilon_+ \right). \nonumber 
\end{align}
The ground states $\ket{\Psi}$ describing the two phases, are now completely characterized as Gaussian states by their Wigner function (\ref{Wigner}), and the corresponding CM (\ref{CM}) and first-moment vector (\ref{R}) are now expressed in terms of the physical parameters $\{\lambda, \omega,\omega_0,N\}$. We notice that the dependence on the size $N$ of the atomic subsystem is contained only in the first-moment vector (\ref{R}). 
\par
\begin{figure}[t]
\includegraphics[width=0.4\textwidth]{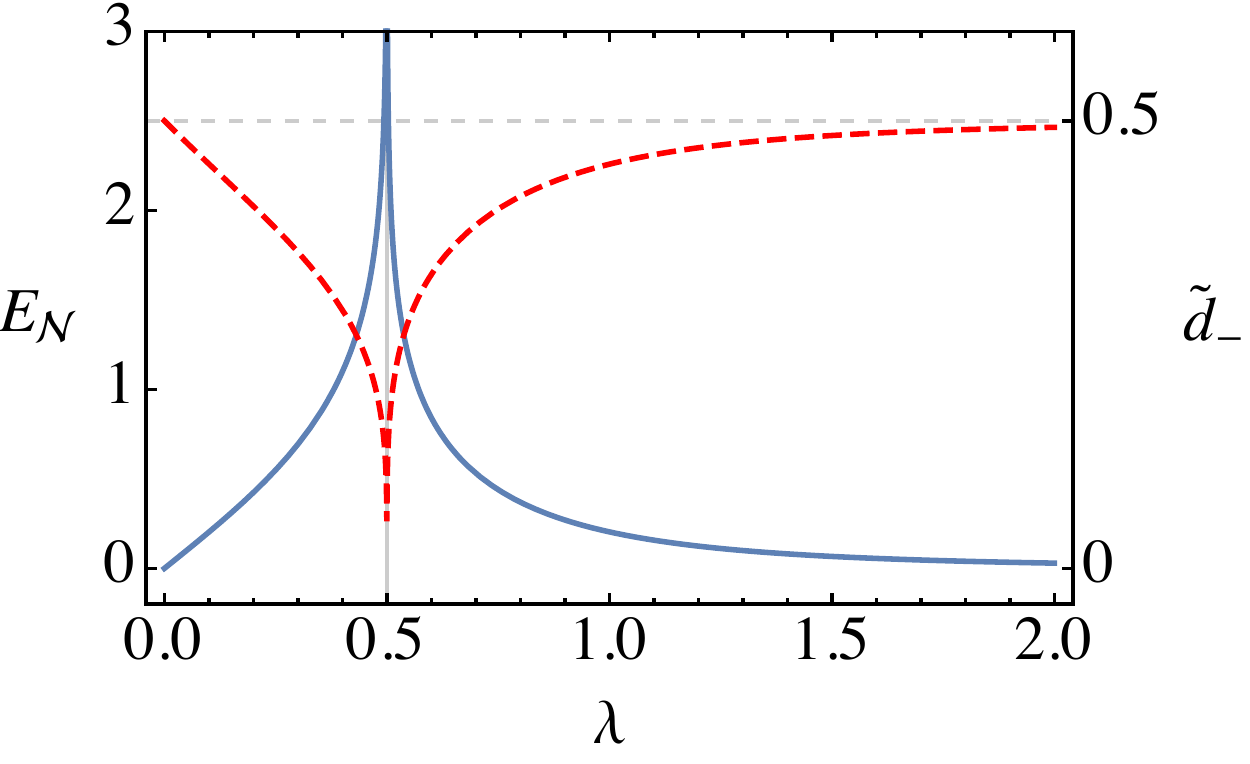}%{scheme.pdf}
%\vspace{-0.2cm}
\caption{(Color online) Plot of the logarithmic negativity $E_{\mathcal{N}}(\lambda)$ (blue solid curve) and of the lowest symplectic eigenvalue $\tilde{d}_-(\lambda)$ (red dashed curve) of the partially transposed CM. The dashed gray line at $\tilde{d}_-=0.5$ represents the threshold of separability, under which the state is entangled. Parameters: $\omega=\omega_0=1$ and $\lambda_c=0.5$ (gray vertical line), in units of $\omega_0$. }
\label{f:GSent}
\end{figure}

In both phases the ground state is a pure Gaussian state ($\mu=1$), with $d_\pm=1/2$, since it has been obtained by a symplectic transformation of the vacuum state $\ket{\psi}$. When the coupling $\lambda$ between the two subsystems gets stronger, the two become increasingly entangled, as witnessed by the logarithmic negativity (\ref{LN}), which quantifies in a monotonic way the violation of ppt-criterion for the separability of a bipartite state. As it is shown in Fig. \ref{f:GSent}, the atomic and radiation subsystems get increasingly entangled as their coupling approaches the critical value $\lambda_c$. The already established result that entanglement enhances the precision of a measurement \cite{DAriano,Giovannetti} will be confirmed in the following, where we will adopt the QET approach to the considered critical system.
%%%
\subsection{QFI and SLD}\label{s:QFI}
Once the ground states in the two phases are known, it is possible to
study the behavior of the QFI, as a function of the coupling parameter
$\lambda$ driving the QPT and the tunable radiation frequency $\omega$,
which sets the critical point $\lambda_c$. We point out that in our
model $\lambda$ and $\omega$ are considered independent on each other,
for the sake of simplicity, but that in some experimental realizations
(see, e.g., Ref.~\cite{Esslinger}) they may be related to the tunable
parameters of an external pumping. \par
\begin{figure}[b]
\includegraphics[width=0.35\textwidth]{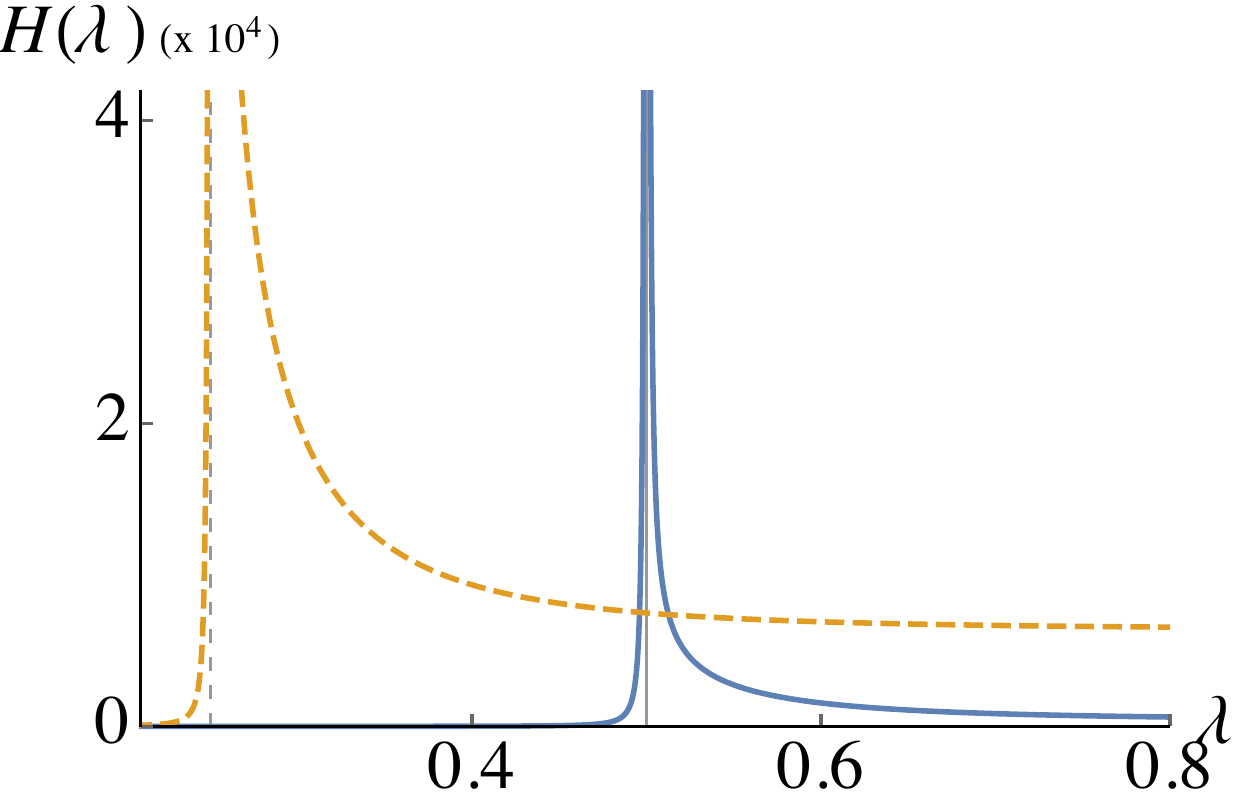}
\caption{(Color online) Plot of the QFI as a function of $\lambda$, where all the quantities are computed in units of $\omega_0$ and $N=100$. Resonance condition: $\omega_0=\omega=1$ (solid blue curve) with $\lambda_c=0.5$. Off-resonance condition: $\omega=0.25$ (dashed orange curve) with $\lambda_c=0.25$. }
\label{f:QFI}
\end{figure}

Referring to Eq.
(\ref{QFI_Gaussian}), it is possible to analytically evaluate the QFI in
the two phases, but we report here only the limiting behaviors in
proximity of the critical value $\lambda_c$. In particular, the leading
term in the series expansion of the QFI approaching the critical
parameter from both the two phases, is  $H(\lambda)\sim
[2\sqrt{2}(\lambda-\lambda_c)]^{-2}$, whereas the main limiting cases
are displayed in Table \ref{t:QFI}.  At the critical point the QFI for
the whole radiation-atoms system diverges with a second-order
singularity, thus highlighting the possibility to estimate the parameter
$\lambda$ (in the ideal thermodynamic limit) with infinite precision. By
tuning $\lambda_c$ with $\omega$, it is possible to obtain the highest
precision for every value of the coupling parameter $\lambda$, as the
behavior of the QFI at $\lambda_c$ is left unvaried (see Fig.
\ref{f:QFI}). We point out that the second term in Eq.
(\ref{QFI_Gaussian}) is non-zero in the superradiant phase, in
particular the QFI behaves in the thermodynamic limit as a linear
increasing function of $N$, with finite-size corrections of the order
$N^{-1/2}$, for every value of the coupling $\lambda$. Nonetheless, at
the critical point $\lambda_c$ the dominant contribution to $H(\lambda)$
is ruled by the coupling parameter (see Table \ref{t:QFI}).
\par
Now we compute the SLD operator in the two phases and analyze the
asymptotic behaviors, with respect to $\lambda$, at the critical point.  In the normal phase the
second term of Eq. (\ref{SLD_Gaussian}) is null, since the amplitudes of
the displacements (\ref{displacement}) are zero.
\begin{table}[t]
\caption{\label{t:QFI}Limiting behaviors of the QFI in the normal and superradiant phases at $\lambda\to\lambda_c^\pm$, $\lambda\to0$ and $\lambda\to\infty$.}\vspace{2mm}
\begin{ruledtabular}
\begin{tabular}{ccc}
& Normal phase & Superradiant phase \\[2mm]
\hline\\[.5mm]
$\lambda\to\lambda_c$ & $\frac{1}{8(\lambda-\lambda_c) ^2} +O\left [\frac{1}{|\lambda-\lambda_c |}\,\right ]$ & $\frac{1}{8(\lambda-\lambda_c) ^2}+O\left [\frac{1}{|\lambda-\lambda_c |}\,\right ]$ \\[3mm]
$\lambda\to 0$ & $\frac{4}{(\omega+\omega_0)^2} + O[\lambda^2]$ & ---  \\[2mm]
$\lambda\to\infty$ & --- & $\frac{4 N}{\omega^2}+O\left [\lambda^{-4}\,\right ]$ \vspace{2mm}
\end{tabular}
\end{ruledtabular}
\end{table}
In both phases $\nu=0$ and the main term of the SLD has the same dependence $|\lambda-\lambda_c|^{-3/2}$, namely 
\begin{equation}\label{SLD1}
\vec{R}^{\,T} (-\dot{\sigma}) \vec{R}\sim \frac{\sqrt[4]{\omega\, \omega_0}}{8\sqrt{2}\sqrt{\omega^2+\omega_0^2}}\frac{1}{|\lambda-\lambda_c|^{3/2}}(\hat{x}_{1}'-\hat{x}_{2}')^2,
\end{equation}
where $\vec{R'}=F_1 \vec{R}$ is the vector of quadratures transformed according to the local squeezing employed in the Hamiltonian diagonalization (\ref{SympTrans}). In the superradiant phase the linear term of the SLD, dependent also on the number of atoms $N$, is constant very close to the critical point, namely 
\begin{equation}\label{SLD2}
\vec{R}^T\vec{\zeta}\sim\sqrt{\frac{ 32 N}{\omega^3\omega_0^2(\omega^2+\omega_0^2)}}\big( \omega_0^2\, \hat{p}_{1}'-\omega^2\, \hat{p}_{2}'\big ),
\end{equation}
in such a way that, ultimately, the SLD diverges at $\lambda_c$ as in Eq. (\ref{SLD1}), but still more slowly than the QFI (see Table \ref{t:QFI} for comparison). Since the SLD is associated to the optimal POVM saturating the quantum Cram\'er-Rao bound (\ref{QCR}), we note that Eq. (\ref{SLD1}) contains a combination of position quadratures relative to both the atomic and radiation subsystems, confirming the highly entangled nature of the two (see Fig. \ref{f:GSent}). 
\par
In the next section we will show that it is still possible to optimally estimate the parameter $\lambda$ around the critical point, by means of locally feasible measurements.
%=-(\dot{\sigma_{11}} q_a^2 +\dot{\sigma_{33}} q_b^2+2\dot{\sigma_{13}} q_a q_b+\dot{\sigma_{22}} p_a^2 +\dot{\sigma_{44}} p_b^2+2\dot{\sigma_{12}} p_a p_b)
\begin{table*}[t]
\caption{\label{t:FIQFI}Limiting behaviors of the FI for homodyne-like detection of both radiation $F_{\hat{x}(\phi)}(\lambda)$ and atomic $F_{\hat{y}(\phi)}(\lambda)$ subsystems (with respect to QFI), in the normal and superradiant phases at $\lambda\to\lambda_c^\pm$, $\lambda\to0$ and $\lambda\to\infty$.}
\vspace{2mm}
\begin{ruledtabular}
\begin{tabular}{ccccc}
 & \multicolumn{2}{c}{Normal phase}&\multicolumn{2}{c}{Superradiant phase}\\ 
  & $\lambda\to 0$ & $\lambda\to\lambda_c^-$ & $\lambda\to\lambda_c^+$ & $\lambda\to\infty$\\[1mm] \hline \\[-.5mm]
$F_{\hat{x}(\phi)}(\lambda)/H(\lambda)$ & $\frac{2[\omega+(\omega+\omega_0)\cos(2\phi)]^2}{\omega^2(\omega+\omega_0)^2}\lambda^2 +O[\lambda^3]$ & $1 + O[\sqrt{|\lambda-\lambda_c|}]$ & $1 + O[\sqrt{|\lambda-\lambda_c|}]$ & $\cos^2\phi + O[\lambda^{-4}]$ \\[3mm]
$F_{\hat{y}(\phi)}(\lambda)/H(\lambda)$ & $\frac{2[\omega_0+(\omega+\omega_0)\cos(2\phi)]^2}{\omega_0^2(\omega+\omega_0)^2}\lambda^2 +O[\lambda^3]$ & $1 + O[\sqrt{|\lambda-\lambda_c|}]$ & $1 + O[\sqrt{|\lambda-\lambda_c|}]$ & $O[\lambda^{-6}]$
\vspace{2mm}
\end{tabular}
\end{ruledtabular}
\end{table*}
\section{Optimal local measurements}\label{s:FI}
The main results of this work are examined in depth in this section and concern the possibility to probe one of the two subsystems (radiation mode or atomic ensemble) with local and handy measurements, in order to retrieve the optimal FI. In particular, we address the two most known and employed optical techniques for measuring and characterizing a single-mode radiation, namely homodyne detection and photon counting. 

\subsection{Homodyne detection}
Since all the information about the radiation mode is encoded in its Wigner function, it is possible to reconstruct the corresponding Gaussian state $\hat{\varrho}$ using the homodyne tomography technique, i.e. repeatedly measuring the field mode quadratures according to the set of observables
\begin{equation}
\hat{x}(\phi)=\frac{\hat{a}{\rm e}^{-\imm\phi}+\hat{a}^\dag{\rm e}^{\imm\phi}}{\sqrt{2}}\equiv \hat{U}^\dag(\phi) \hat{x} \,\hat{U}(\phi),
\end{equation}
where $\hat{U}(\phi)\equiv {\rm e}^{-\imm \phi\, \hat{a}^\dag \hat{a}}$ is a phase-shift operator. The probability distribution of the possible outcomes of a quadrature-measurement $p_{x(\phi)}=\meanvalue{x}{\hat{U}(\phi) \hat{\varrho}\, \hat{U}^\dag(\phi)}{x}$, corresponds to the marginal distribution
\begin{equation}\label{Px_Wigner}
p_{x(\phi)}=\int_{\mathbb{R}} \text{d}p\, W[\hat{\varrho}](x \cos\phi - p\sin\phi, x\sin\phi + p\cos\phi),
\end{equation}
where the Wigner function $W[\hat{\varrho}](x,p)$ of the reduced state of the radiation mode (see Sec.~\ref{Gaussian}) $\hat{\varrho}=\text{Tr}_2[\ketbra{\Psi}{\Psi}]$ is Gaussian with second and first moments given by 
\begin{align}
\sigma&=\begin{pmatrix}
\sigma_{11} & 0 \\
0 & \sigma_{22} 
\end{pmatrix} \label{CM1}\\[3mm]
\langle\vec{R}\rangle &= (\alpha\sqrt{2 N},0)^T \label{R1}.
\end{align}
\begin{figure}[b]
\includegraphics[width=0.23\textwidth]{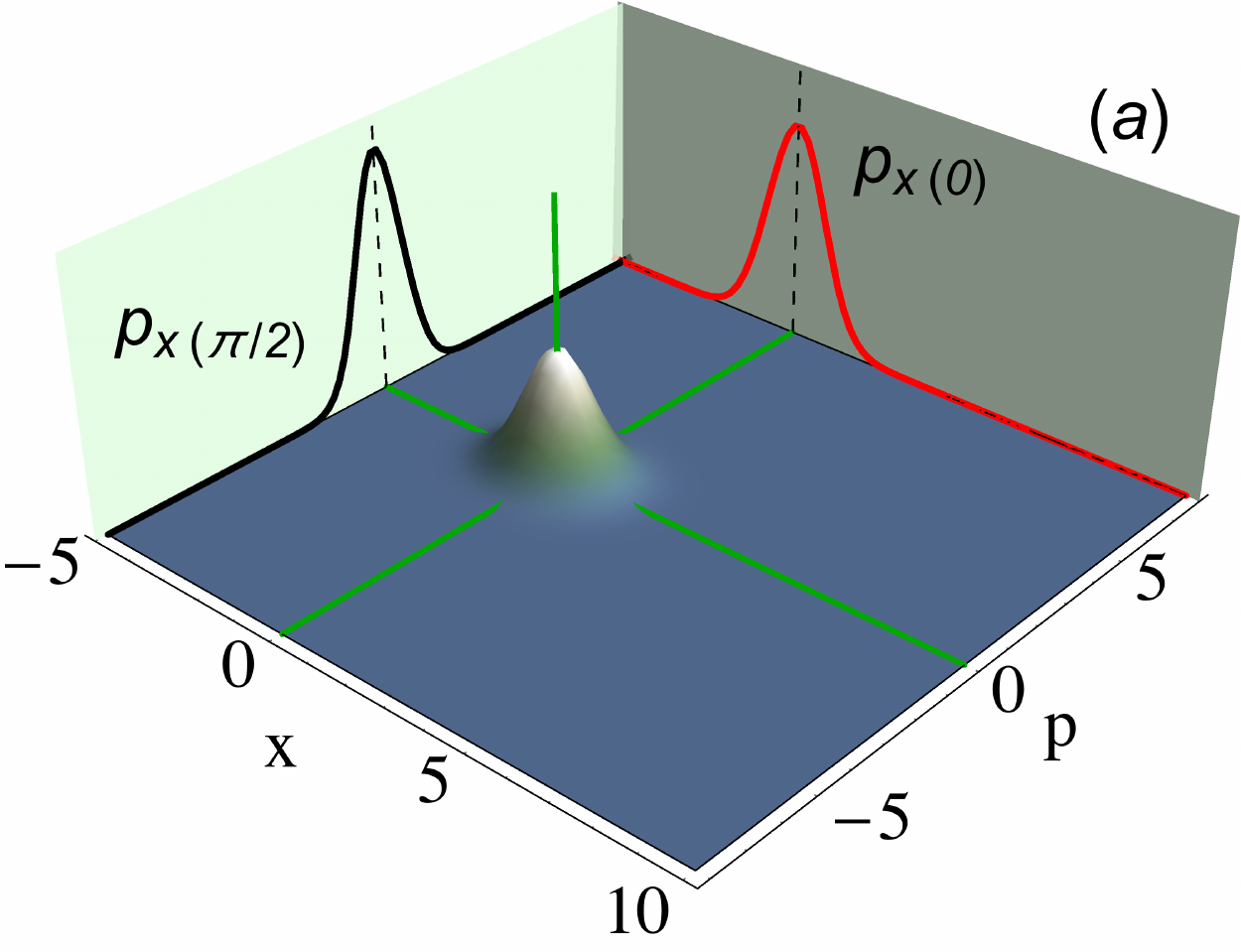} \includegraphics[width=0.23\textwidth]{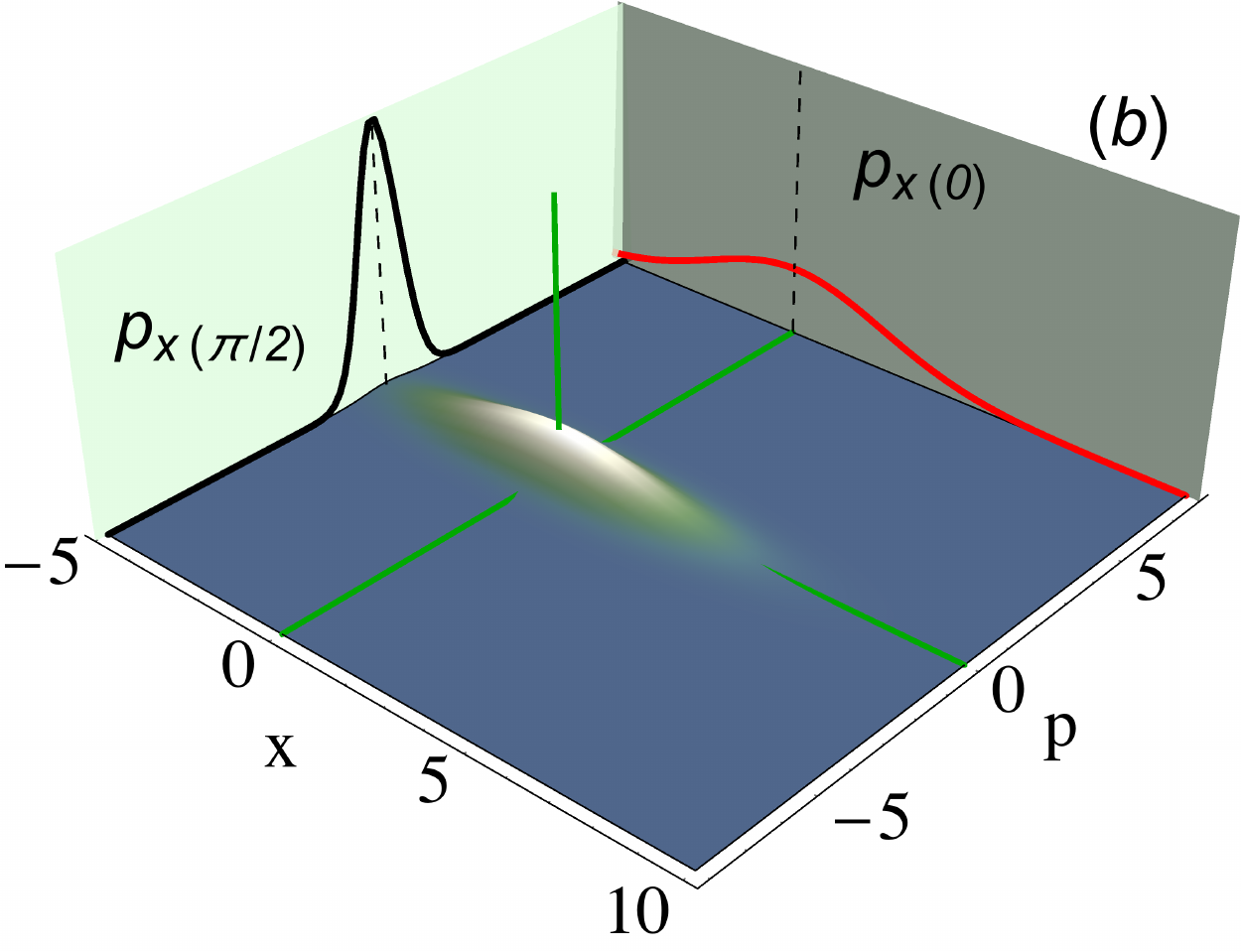}\\
\includegraphics[width=0.23\textwidth]{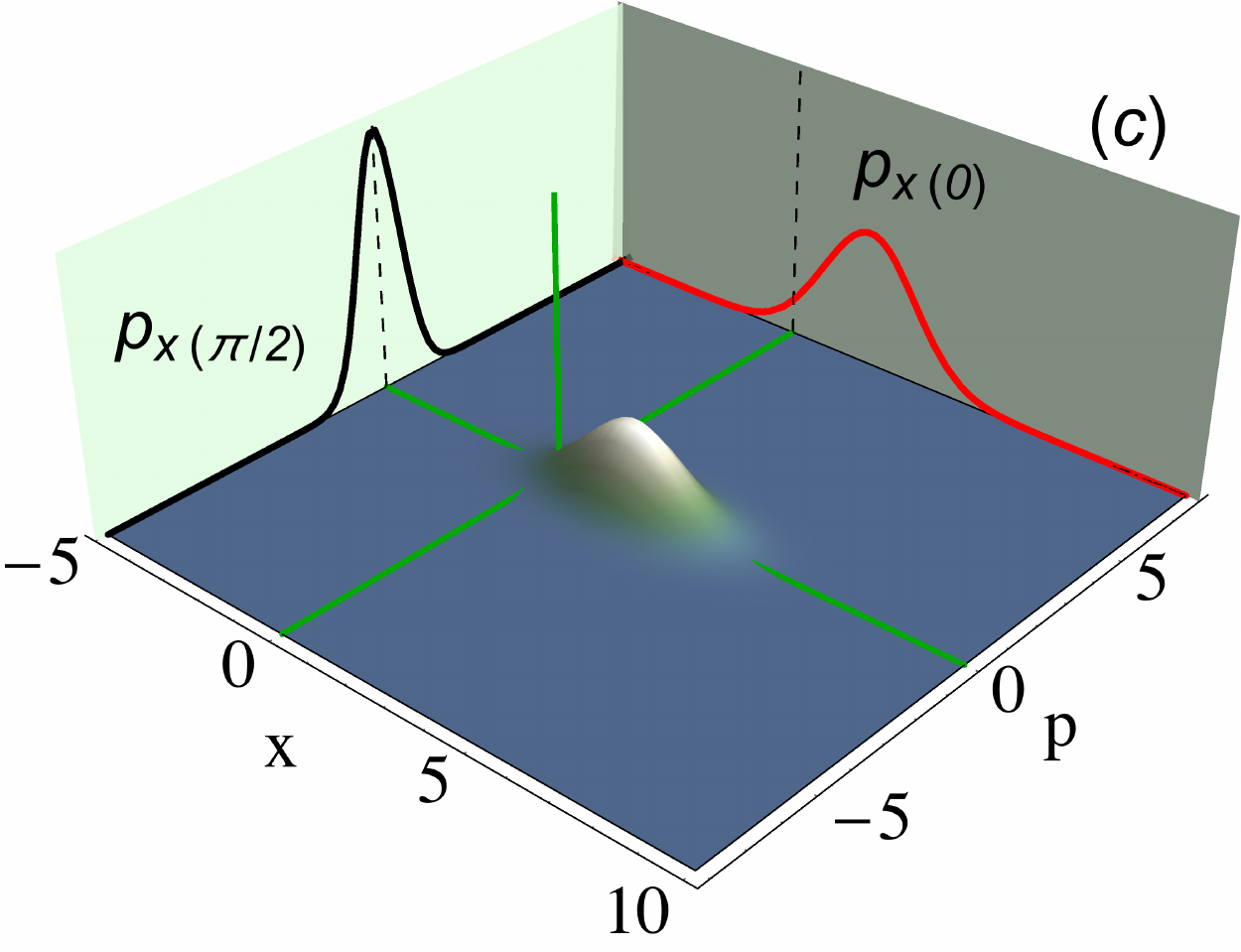} \includegraphics[width=0.23\textwidth]{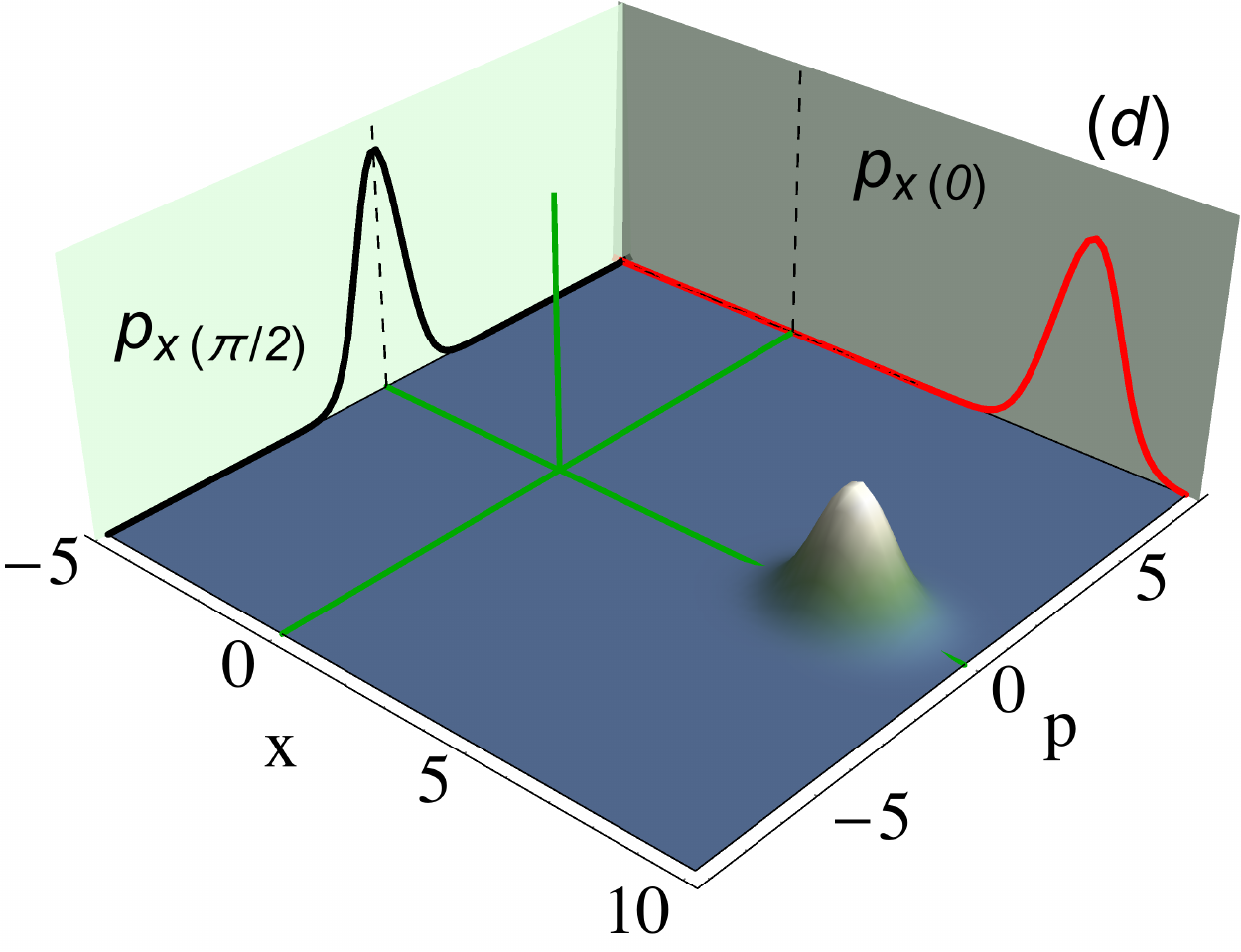}
\caption{(Color online) Wigner function $W[\hat{\varrho}](x,p)$ of the radiation mode state at different values of $\lambda=0.3$ (a), $\lambda=0.499$ (b), $\lambda=0.6$ (c) and $\lambda=1.5$ (d). Marginal distributions, corresponding to the probability for the position and momentum quadratures, respectively $p_{x(0)}$ and $p_{x(\pi/2)}$, are also shown. The values of the chosen parameters are $\omega_0=\omega=1$ and $\lambda_c=0.5$ (in units of $\omega_0$), in the superradiant phase $N=100$.}
\label{f:Wigner}
\end{figure}
In Fig.~\ref{f:Wigner} we show the Wigner function associated to the radiation subsystem, together with the marginal distributions corresponding to homodyne measurements of the position $\hat{x}(0)$ and momentum $\hat{x}(\pi/2)$. From the sequence of frames at different values of $\lambda$, the QPT is evident, where the field mode essentially undergoes a strong squeezing around $\lambda_c$ and then a displacement for $\lambda>\lambda_c$.
\par
We now evaluate the FI associated to the homodyne measurement probing the Gaussian ground state of the radiation mode, as a function of the parameter $\lambda$ driving the QPT. Since the probability distribution (\ref{Px_Wigner}) has Gaussian form with mean value $\langle \hat{x}(\phi) \rangle=\cos\phi\langle \hat{x}(0) \rangle $ and variance $\sigma(\phi)=\cos^2\phi\,\sigma_{11}+\sin^2\phi\,\sigma_{22}$, it is straightforward to derive a general expression for the FI (\ref{FI}) valid for both the normal and superradiant phases
\begin{equation}\label{FIx}
F_{\hat{x}(\phi)}(\lambda)=\frac{2\sigma(\phi) \langle \hat{x}(\phi) \rangle^2+\dot{\sigma}^2(\phi)}{2 \sigma^2(\phi)}.
\end{equation}
\begin{figure}[t]
\includegraphics[width=0.35\textwidth]{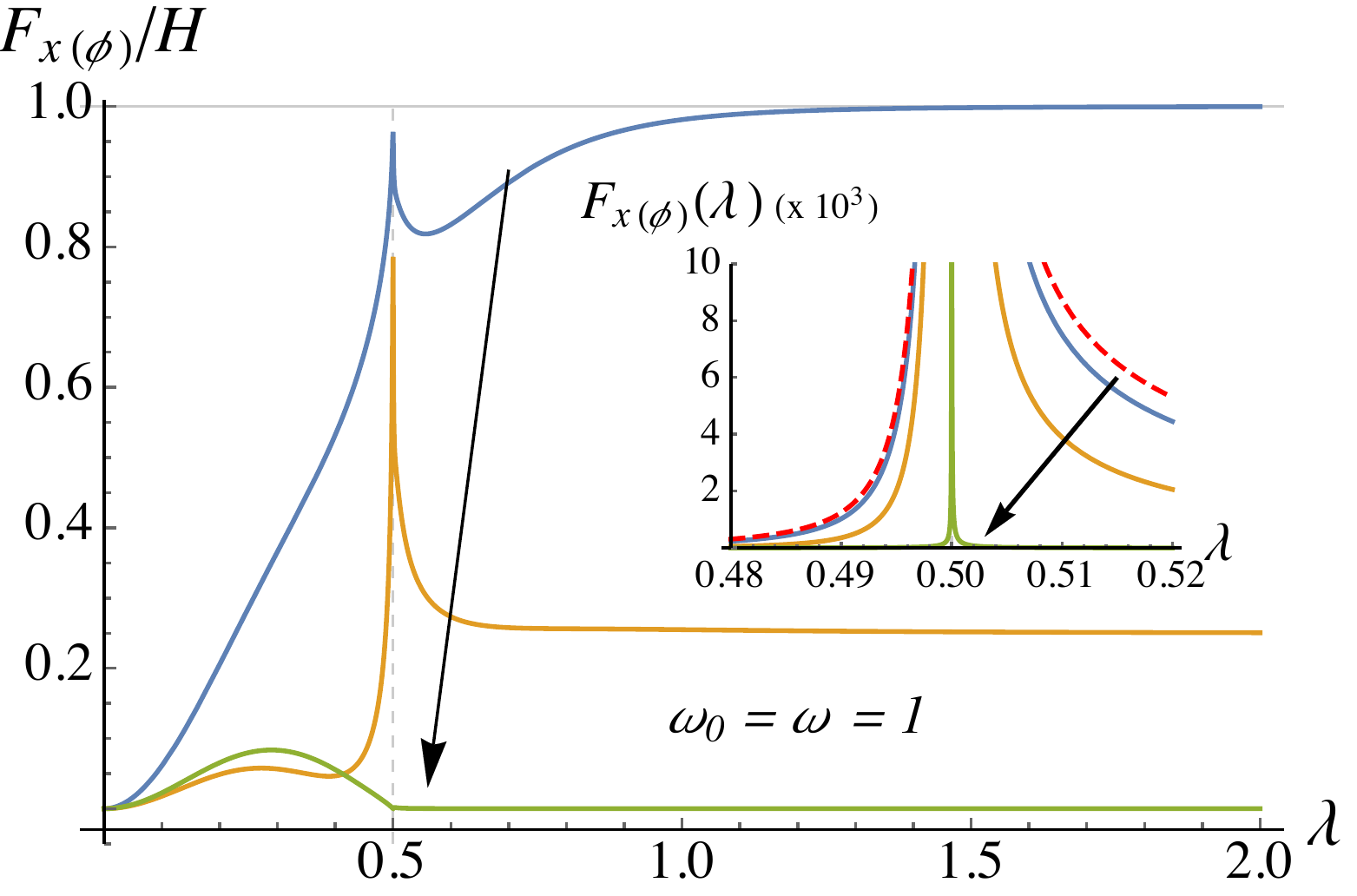}
\includegraphics[width=0.35\textwidth]{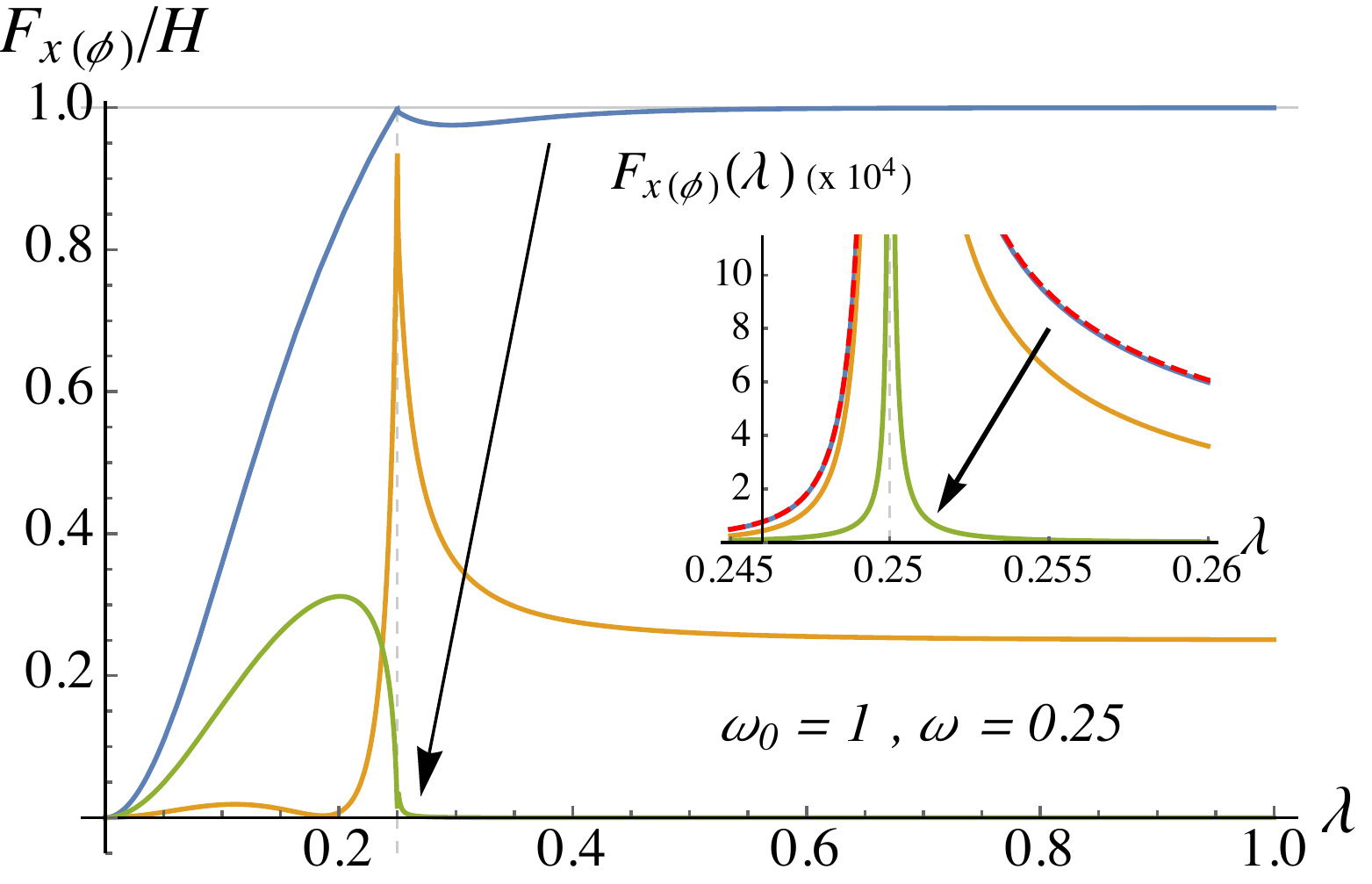}
\caption{(Color online) Plot of the ratio between FI for homodyne detection and QFI, as a function of $\lambda$. The insets show the behavior of the FI (solid curves) and the QFI (dashed curve), both diverging at the critical parameter $\lambda_c$. The arrows indicate the increasing values of the quadrature angle $\phi = 0,\pi/3,\pi/2$ (solid curves). Upper panel: resonance condition with $\omega_0=\omega=1$ and $\lambda_c=0.5$. Lower panel: off-resonance condition with $\omega_0=1$, $\omega=0.25$ and $\lambda_c=0.25$. In both cases the set of parameters is in units of $\omega_0$ and $N=100$.}
\label{f:FIQFI}
\end{figure}
The scaling behaviors of the FI, compared to the QFI, are listed in Table \ref{t:FIQFI} for both the normal and superradiant phases. It is remarkable that a measurement only on a part of the system, namely the radiation mode subsystem, provides the optimal value of the FI in proximity of the critical point. Homodyne detection results to be an optimal local measurement, easily feasible with standard optical techniques, able to provide the best performances in parameter estimation and to capture the quantum criticality. In Fig. \ref{f:FIQFI} we show that for different values of the angle of the measured quadrature $\hat{x}(\phi)$, at the critical point $\lambda\to\lambda_c^\pm$ FI diverges with the very same scaling behavior of QFI, saturating the quantum Cram\'er-Rao bound (\ref{QCR}). The only exception, which do not invalidate the homodyne measurement, is that exactly at $\phi=\pi/2$ the FI is no longer optimal at $\lambda_c$, even though its diverging character (see the insets in Fig. \ref{f:FIQFI}) represents a high precision measurement according to the classical Cram\'er-Rao bound (\ref{CR}). 
Besides, we point out that the FI, in the superradiant phase and  in the
thermodynamic limit, scales as a linear function of $N$, with
finite-size corrections of the order $N^{-1}$. Thus, the ratio between
QFI and FI plotted in Fig.~\ref{f:FIQFI} is essentially independent on
$N$, for every value of the coupling $\lambda$. 
\par
Analogously, a homodyne-like detection of the atomic subspace,
corresponding to measure the generic component $\hat{J}(\phi)\equiv
\hat{J}_x \cos\phi +\hat{J}_y \sin\phi$ of the collective atomic spin in
the $\{x,y\}$-plane, results to be optimal at the critical coupling
$\lambda_c$. The only differences are: (i) in the limit $\lambda\to 0$,
the atomic and radiation frequencies, $\omega_0$ and $\omega$, are
interchanged and (ii) in the limit $\lambda\to\infty$, the FI goes to
zero (see Table \ref{t:FIQFI}).  \par
Interestingly, the electromagnetic field quadratures appear in the
limiting expression for the SLD (\ref{SLD1}), thus confirming the
optimal character of the chosen homodyne-type detection employed to
probe just one of the two subsystems.

\subsection{Photon counting}

Another typical observable used to probe the electromagnetic field is the photon number operator
\begin{equation}\label{Obs_N}\begin{split}
&\hat{N}_1\equiv\hat{a}_1^\dag \hat{a}_1=\sum_{n=0}^{\infty}n\ketbra{n}{n} \\
&\sum_{n=0}^{\infty} n \,p(n)=\text{Tr}[\hat{\varrho}\, \hat{N}_1],
\end{split}\end{equation}
\begin{figure}[b]
\includegraphics[width=0.35\textwidth]{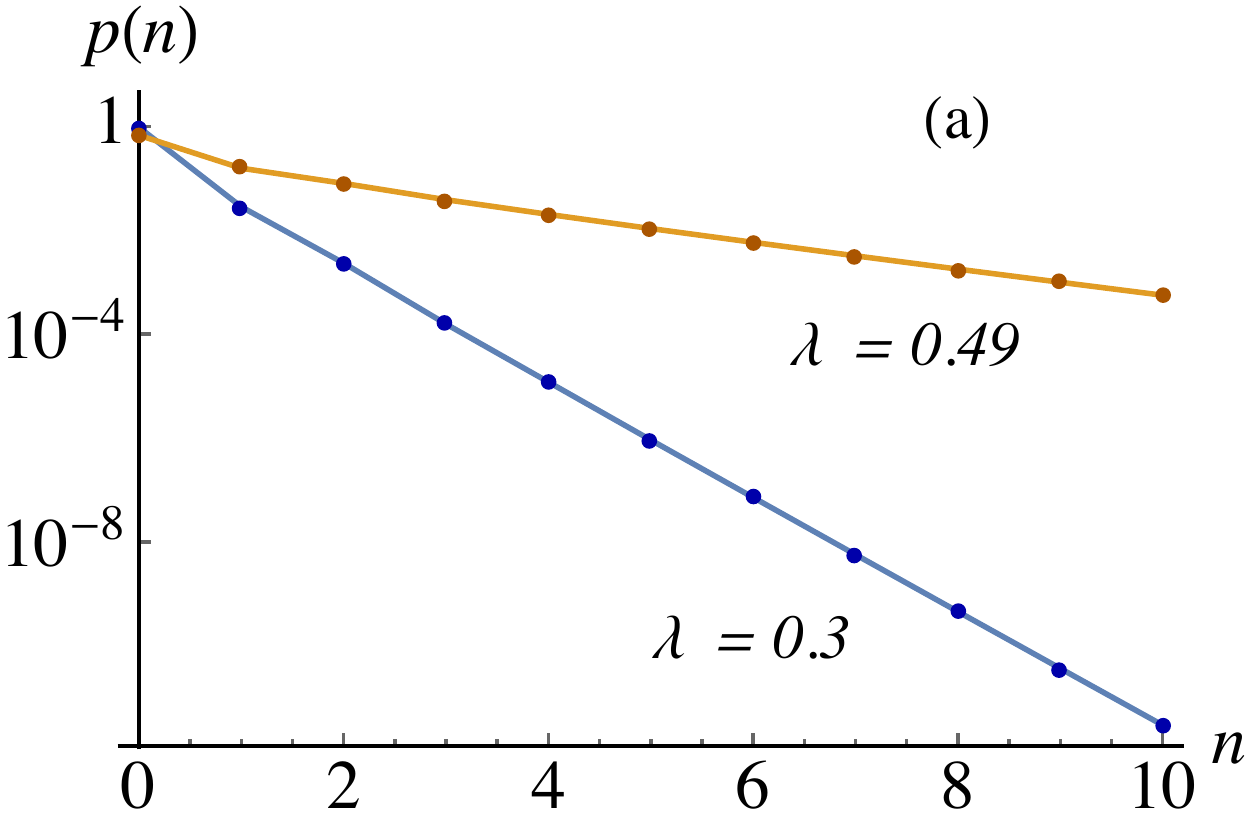}\\[2mm]
\includegraphics[width=0.35\textwidth]{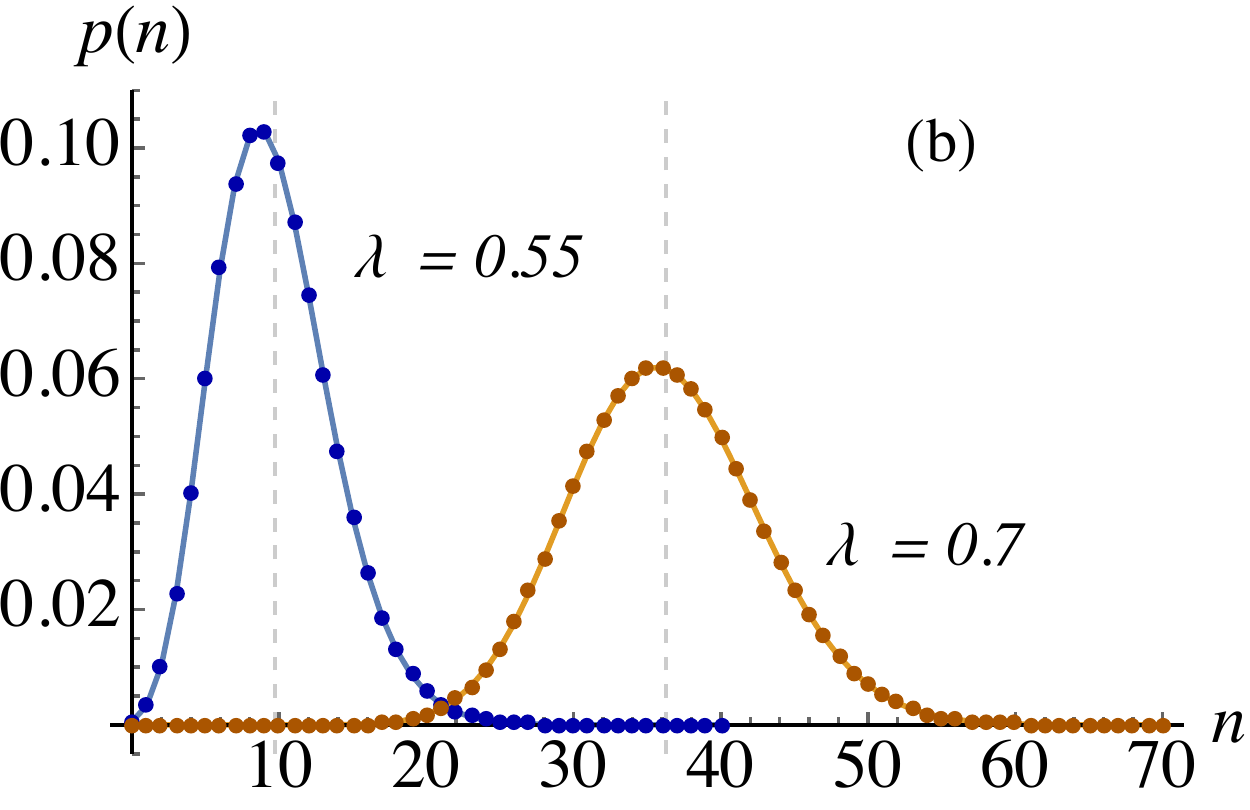}
\caption{(Color online) Logarithmic plot of photon number probability distributions (\ref{pn}) of the radiation ground state in the normal phase (a), with $\omega_0=\omega=1$ and $\lambda=0.3,0.49$. Photon number probability distributions (\ref{pn}) of the radiation ground state in the superradiant phase (b), with $\omega_0=\omega=1$, $N=100$ and $\lambda=0.55,0.7$. The mean values of the distributions, Eq. (\ref{Nave}), are specified with dashed vertical lines. }
\label{f:pn}
\end{figure}
where $p(n)=\meanvalue{n}{\hat{\varrho}}{n}$ is the probability to detect a photon in the Fock state $\ket{n}$. Photon counters capable of discriminating among the number of incoming photons are commonly employed in quantum optical experiments \cite{Esslinger,Wittman,Bondani}. As we mentioned in Sec.~\ref{Gaussian}, the partial trace of a Gaussian bipartite state, is a single-mode Gaussian state which can be cast in the general form of a DSTS  $\hat{\varrho}=\hat{D}(\gamma)\hat{S}(r) \hat{\nu}_\text{th}(\bar{n}) \hat{S}^\dag (r)\hat{D}^\dag (\gamma)$. The analytic and general expression for the photon number probabilities \cite{Marian}, applied to the state of the radiation subsystem with CM and first-moment vector given by Eq. (\ref{CM1}) and Eq. (\ref{R1}), respectively, reads
\begin{equation}\label{pn}\begin{split}
&p(n) = R_{00} (-1)^{n} 2^{-2n} (\tilde{A}+|\tilde{B}|)^{n} \times \\[1mm]
&\times\sum_{k=0}^{n} \frac{H_{2 k} (0) H_{2 n - 2 k} \left (  \imm\,\tilde{C}\left [ \tilde{A} + |\tilde{B}| \right ]^{-\frac12} \right )}{k! (n - k)!}  \left [ \frac{\tilde{A}-|\tilde{B}|}{\tilde{A}+|\tilde{B}|}\right ]^k ,
\end{split}\end{equation}
where $H_{m} (x)$ are Hermite polynomials. All the quantities appearing in Eq. (\ref{pn}) depend only on first- and second-moments as follows:
\begin{equation}\begin{split}
R_{00}& =\frac{2\exp\{ -\frac{\langle \hat{x}_1 \rangle^2}{1+2 \sigma_{11}} \}}{\sqrt{(1+2 \sigma_{11})(1+2 \sigma_{22})}} \\
\tilde{A}& = \frac{4 \sigma_{11} \sigma_{22}-1}{(1+2 \sigma_{11})(1+2 \sigma_{22})} \\
\tilde{B}& = \frac{ 2(\sigma_{22} - \sigma_{11})}{(1+2 \sigma_{11})(1+2 \sigma_{22})} \\
\tilde{C}& = \frac{\sqrt{2}\,\langle \hat{x}_1 \rangle}{1+2 \sigma_{11}}.
\end{split}\end{equation}
\begin{figure}[t]
\includegraphics[width=0.35\textwidth]{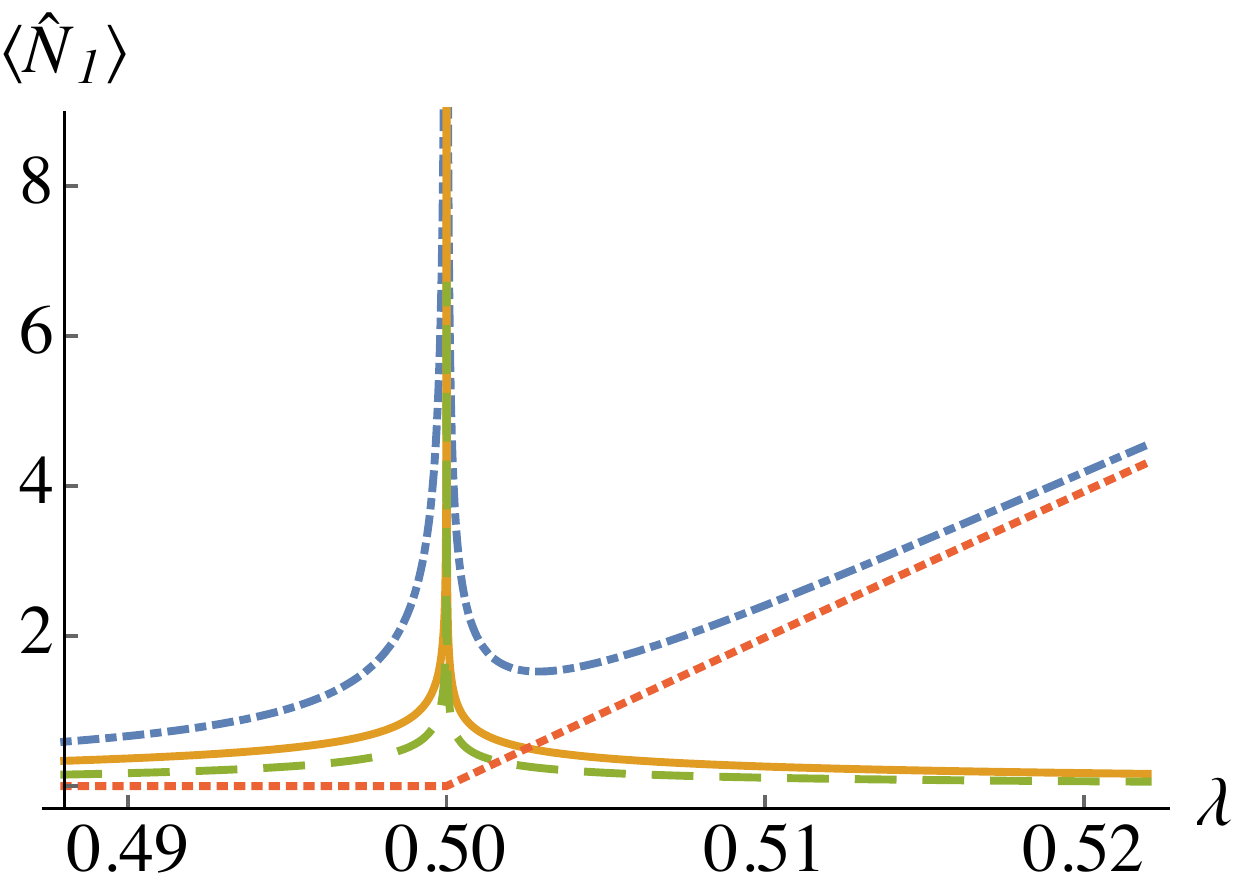}
\caption{(Color online) Plot of the mean energy of the radiation mode subsystem (dot-dashed curve) as a function of the coupling parameter $\lambda$. Three contributions to $\langle \hat{N}_1 \rangle$ are showed: mean thermal photons $\bar{n}$ (solid curve), mean squeezed photons $n_s$ (dashed curve) and mean coherent energy $|\alpha|^2 N$ (dotted curve).}
\label{f:Nave}
\end{figure}
In Fig. \ref{f:pn} we plot the probability distributions for the photon number characterizing the ground states of the two phases. In the normal phase, the reduced ground state for the radiation subsystem is a squeezed thermal state with typical photon number distribution peaked in $n=0$, whereas in the superradiant phase it acquires macroscopic occupation due to the non-zero displacement amplitude (\ref{displacement}). 
The general expression of the mean photon number of a generic single-mode Gaussian state in the DSTS form is 
\begin{equation}\label{Nave}
\langle \hat{N}_1 \rangle = n_s +\bar{n} (1+2 n_s)+|\gamma|^2.
\end{equation}
It is possible to identify an intensive contribution to $\langle \hat{N}_1 \rangle$ given by the mean number of thermal photons $\bar{n}=\sqrt{\sigma_{11}\sigma_{22}}-1/2$ and the fraction of squeezed photons $n_s=\sinh^2 r$, with $r={\rm Log} (\sqrt[4]{\sigma_{11}/\sigma_{22}})$. The extensive contribution is provided by the amplitude of displacement $\gamma=\alpha\sqrt{N}$, depending on the number of atoms. As plotted in Fig.~\ref{f:Nave}, it is evident how, in proximity of the phase transition, the mean photon number dramatically increases due to a strong degree of squeezing and a high thermal component. Only in the superradiant phase the extensive contribution $|\alpha|^2 N$ dominates far away of the critical parameter, due to an increasing coherent state component (see also Fig.~\ref{f:Wigner}(d)). We point out that even in the thermodynamic limit, although in the normal phase the extensive contribution is not present, in the proximity of the critical point a non-negligible fraction of squeezed thermal photons should be measured by a photodetector.
\par
The FI information associated to the observable (\ref{Obs_N}) is given by Eq. (\ref{FI}) expressed in discrete form
\begin{equation}\label{FIN}
F_{\hat{N}_1}(\lambda)=\sum_{n=0}^\infty \frac{\left [\partial_\lambda p(n)\right ]^2}{p(n)}.
\end{equation}
\begin{figure}[t]
\includegraphics[width=0.35\textwidth]{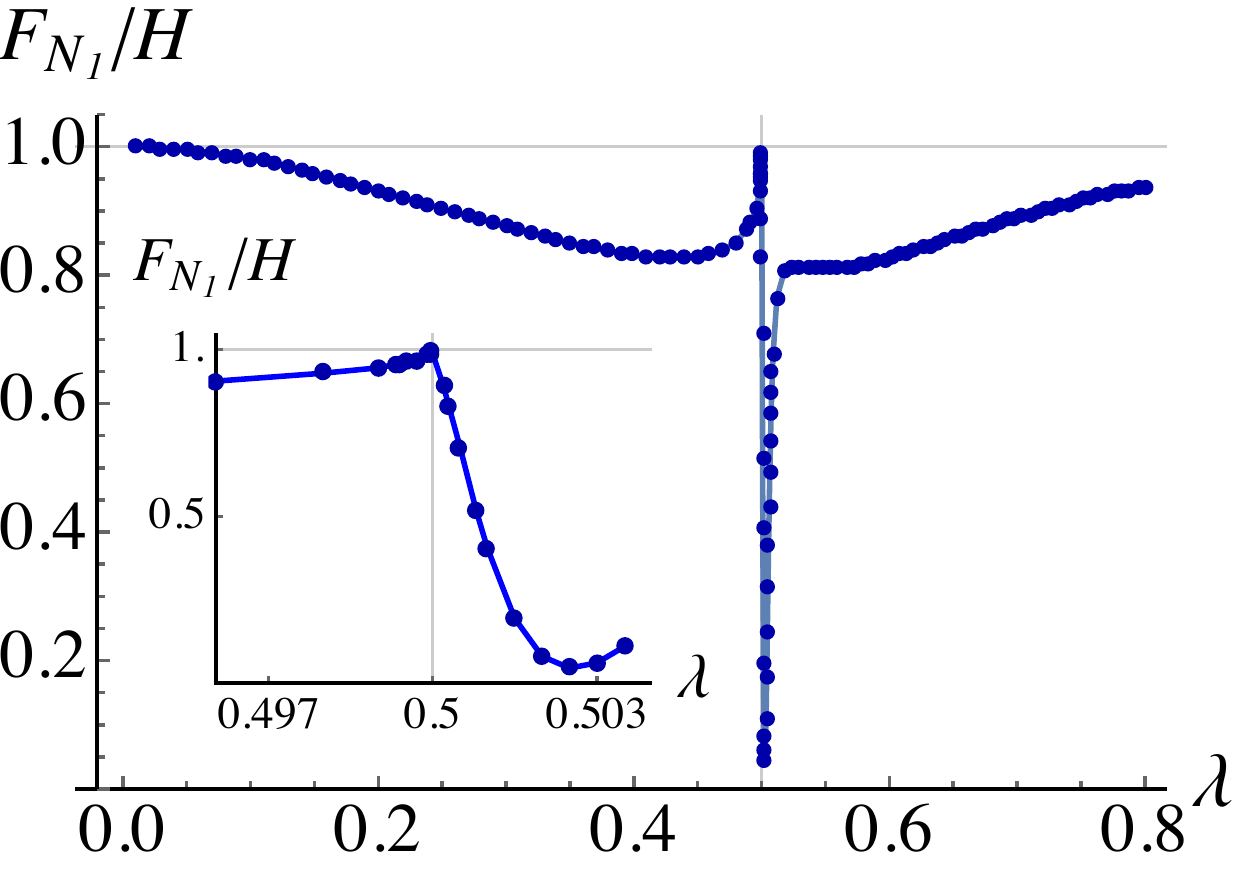}
\caption{(Color online) Plot of the ratio between FI for a photon-count measurement and QFI, as a function of $\lambda$. The inset shows a magnification around the critical parameter $\lambda_c$, showing in a clearer way that the observable $\hat{N}_1$ is optimal. The values of the parameters (in units of $\omega_0$) are $\omega_0=\omega=1$, $\lambda_c=0.5$ and $N=100$ in the superradiant phase.}
\label{f:FIQFI_Nave}
\end{figure}
In Fig. \ref{f:FIQFI_Nave} we show the behavior of the FI associated to a photon-count measurement compared to the QFI. Even though numerical simulations necessarily imply a cut-off value of the dimensionality of the Fock space in evaluating the series in Eq. (\ref{FIN}), making the numerical calculations awkward around $\lambda_c$, it is evident that the observable $\hat{N}_1$ tends to be optimal at the critical coupling.  We can, thus, strengthen our main result, according to which optimal parameter estimation around the region of criticality can be achieved even by probing only a part of the composite system.

\section{Conclusions}
We have analyzed the superradiant QPT occurring in the 
Dicke model in terms of Gaussian ground states with the 
help of the symplectic formalism. In this framework, we have 
addressed the problem of estimating the coupling parameter, investigating
whether and to which extent criticality is a resource to enhance
precision. In particular, we have obtained analytic expressions and limiting
behaviors for the QFI, showing explicitly its divergence at critical
point. Upon tuning the radiation frequency we may also tune the 
critical region and, in
turn, achieve optimal estimation for any value of the radiation-atoms
coupling. 
\par 
Besides, we studied two feasible measurements to be performed only onto
a part of the whole bipartite system, homodyne-like detection and photon
counting. The remarkable result is that by probing just one of the two
subsystems, namely the radiation mode or the atomic ensemble, it is
possible to achieve the optimal estimation imposed by the quantum
Cram\'er-Rao bound. Notice that this is a relevant feature of the system, 
in view of its strongly interacting nature and of the high degree of 
entanglement of the two subsystems at the critical point. 
The possibility of probing  the system accessing only the radiation 
part is of course a remarkable feature for practical applications.
\par
Motivated by relevant and fruitful experimental interests, recently
arisen in connection to the realization of exotic matter phases, we
believe that a quantum estimation approach, as the one outlined in this
work, can be profitably employed in quantum critical systems. The gain
is twofold, since (i) criticality is a resource for the estimation of
unaccessible Hamiltonian parameters and (ii) the search for optimal
observable providing high-precision measurements allows a fine-tuning
detection of the QPT itself. The analysis may be also extended to finite
temperature and to systems at thermal equilibrium. Work along these
lines is in progress and results will be reported elsewhere.
\begin{acknowledgments}
This work has been supported by EU through the Collaborative Projects
QuProCS (Grant Agreement 641277) and by UniMI through the H2020
Transition Grant 14-6-3008000-625.  \end{acknowledgments}
% Create the reference section using BibTeX:

\end{document}